\documentclass[conference]{IEEEtran}
\IEEEoverridecommandlockouts

\usepackage{cite}
\usepackage{amsmath,amssymb,amsfonts}
\usepackage{algorithmic}
\usepackage{graphicx}
\usepackage{textcomp}
\usepackage{xcolor}

\usepackage{tablefootnote} 
\usepackage{placeins} 

\DeclareMathOperator*{\argmin}{arg\,min}

\def\BibTeX{{\rm B\kern-.05em{\sc i\kern-.025em b}\kern-.08em
    T\kern-.1667em\lower.7ex\hbox{E}\kern-.125emX}}
\begin{document}

\title{Unitary Spreading for Robust LPI/AJ TRANSEC in CPM Systems}

\author{
    \IEEEauthorblockN{Tucker Hathaway and Daniel Chew}
    \IEEEauthorblockA{ Rampart Communications}\
                        Linthicum Heights, MD \\
                        USA \\
    Email: \{thathaway, dchew\}@rampartcommunications.com 
    }

\maketitle

\begin{abstract}
Adversarial feature extraction and blocking jamming threaten tactical CPM links. This paper presents a unitary spreading-based Transmission Security (TRANSEC) enhancement to obscure physical-layer signatures and improve anti-jamming (AJ) resilience. The enhancement can be used to augment existing techniques. The enhancement preserves the constant-envelope (0 dB PAPR) nature of CPM, ensuring compatibility with high-efficiency tactical amplifiers. Analysis of symbol distributions, spectra, and cyclostationary features demonstrates that the technique masks inherent signatures, preventing modulation classification. We leverage convex optimization to recover symbols under blocking jamming, reducing uncoded BER from 6.25\% to 0.04\%. Finally, we characterize the engineering trade-offs between security, bandwidth, and BER. 
\end{abstract}




\section{Introduction}
Securing wireless communications in hostile and contested operational environments requires adherence to TRANSEC principles. While traditional data encryption methods are highly effective at securing the information payload, they inherently leave distinct physical-layer signatures exposed. This creates a critical vulnerability: adversaries can bypass data cryptography entirely by analyzing the waveform's structure or attacking the signal to disrupt reception.

Malicious actors increasingly rely on advanced signal processing techniques to exploit these physical-layer signatures. Because engineered communication signals exhibit hidden periodicities (such as carrier frequencies, symbol rates, and coding schemes), cyclostationary analysis allows adversaries to detect weak transmissions buried in noise, reverse-engineer proprietary waveforms, and classify specific modulation types. Doing so allows an adversary to categorize emitters, collect data symbols for analysis and decryption, and analyze traffic patterns.

To address the gap between payload encryption and physical-layer vulnerability, we investigate a novel TRANSEC enhancement designed to obscure the fundamental properties of CPM signals. In this paper, we detail the implementation of this TRANSEC enhancement and rigorously evaluate its effectiveness using CPM waveforms of varying modulation orders. We provide a comprehensive analysis of the system's resilience to feature extraction by evaluating symbol distributions, spectral densities, occupied bandwidths, and cyclostationary features. Additionally, we analyze the enhancement's resilience to blocking jamming. Furthermore, because physical-layer obfuscation inherently impacts communication efficiency, we systematically evaluate the engineering trade-offs required to balance enhanced TRANSEC capabilities against occupied bandwidth constraints and bit error rate (BER) penalties.

\subsection{TRANSEC, LPI, and AJ}
TRANSEC, Low Probability of Interception (LPI), and Anti-Jam (AJ) are formally defined in \cite{CNSSI4009_2022} as follows:
TRANSEC: "\textit{Measures (security controls) applied to transmissions in order to prevent interception, disruption of reception, communications deception, and/or derivation of intelligence by analysis of transmission characteristics such as signal parameters or message externals.}"
LPI: "\textit{Result of  measures used to resist attempts by adversaries to analyze the  parameters of a transmission to determine if it is a signal of interest.}"
AJ: "\textit{The result of measures to resist attempts to interfere with communications reception.}"

TRANSEC is enabled through measures such as LPI to prevent signal analysis, and AJ to resist the disruption of reception. We will employ unitary spreading to obfuscate waveform features as a means to frustrate adversary attempts at parameterizing the signal. That same unitary spreading will provide a measure of immunity to jamming attacks.

\subsection{Cyclostationary Features}
The exploitation of wireless communications via cyclostationary feature extraction is extensively documented in the literature \cite{GARDNER2006639, spectrum_sense_cyclo}. For example, cyclostationary features can be used to extract critical signal characteristics like symbol rates and modulation types \cite{fehske2005}. If two radio links employ different modulation orders, cyclostationary analysis can easily distinguish between them. These extracted characteristics facilitate waveform reverse-engineering, emitter identification, traffic pattern analysis, and the classification of emitter groups. Therefore, anonymizing and mitigating these features is a critical TRANSEC priority.

\subsection{Unitary Spreading}
Unitary spreading (sometimes called unitary spreading transforms or linear constellation precoding) is the process of applying a unitary transform to a block of modulated symbols. The result is that the symbols are now part of a rotated lattice. This idea has been explored in numerous applications, one of the most well cited being \cite{Boutros1998}. In that reference, a block of QAM is rotated as a means to achieve signal space diversity and improve performance in fading channels. Signal space diversity through constellation rotation is employed in DVB-T2 \cite{Nour2008} \cite{Yang2015}. The carrier offset sensitivities of unitary transforms described in \cite{Kalbat2020} are sidestepped by using CPM waveforms with noncoherent derivative of arctan demodulation.

Unitary spreading is performed at the symbol rate, thus it does not expand the bandwidth of the symbols. The rotated symbols are not spread across a large bandwidth as with other spreading techniques. Instead, the energy of each symbol in a frame is spread over that whole frame. Because the output is a sum of symbols which are independent and identically distributed random variables, the output of this unitary spreading approaches a Gaussian distribution. Unitary spreading can be used with other forms of spreading to augment them.

We further consider the impact of blocking jamming, which saturates the receiver front-end and blanks the post-demodulated data. The existing literature contains examples of blanking samples within a unitary transform frame and explores the impact \cite{Zhidkov2008} \cite{Brandes2009} \cite{Epple2017}. In this work, we explore the effect of blanking on the unitary transformed frame and introduce a box-relaxation convex optimization method \cite{Thrampoulidis2018} to recover the underlying symbols. This method is tunable, scales linearly with the desired number of iterations, and offers a computationally efficient alternative to prohibitively complex maximum likelihood methods and sphere decoding. In this work, the convex method demonstrates a decisive improvement on uncoded data. This method can augment forward error correction (FEC) schemes.

\section{System Model}
\label{sec:system_model}

This work will explore two modulation schemes in order to demonstrate and quantify the benefits of the TRANSEC enhancement. Those two schemes are a 1-bit-per-symbol case (MSK) and a 2-bit-per-symbol case (16-PSK/FM). 

\subsection{Transmitter}
\begin{figure}[!htbp]
    \centering
    \includegraphics[width=1\linewidth]{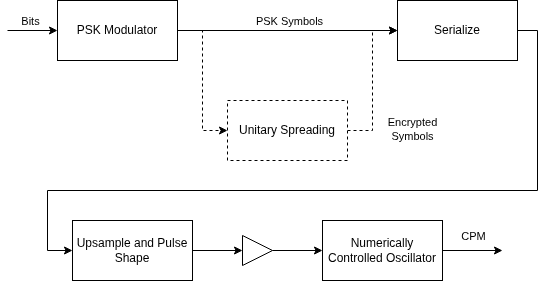}
    \caption{Transmitter System Diagram}
    \label{fig:tx_sys_diagram}
\end{figure}

The flow of the transmitter is shown in Fig. \ref{fig:tx_sys_diagram}. The transmitter will create a digital modulating signal, including pulse shaping, and use that modulating signal to drive a numerically controlled oscillator. 
Data bits are encoded into symbols at the PSK modulator. For MSK, this PSK modulator produces a QPSK constellation. The in-phase and quadrature-phase components of that QPSK will be serialized, thus producing a BPSK sequence of +1s and -1s. Higher-order modulations will exhibit more levels. Fig. \ref{fig:tx_sys_diagram} shows a unitary spread block in dashed lines. When TRANSEC is enabled, the PSK symbols are put through a unitary transform prior to serialization.

\begin{figure}[!htbp]
    \centering
    \includegraphics[width=1\linewidth]{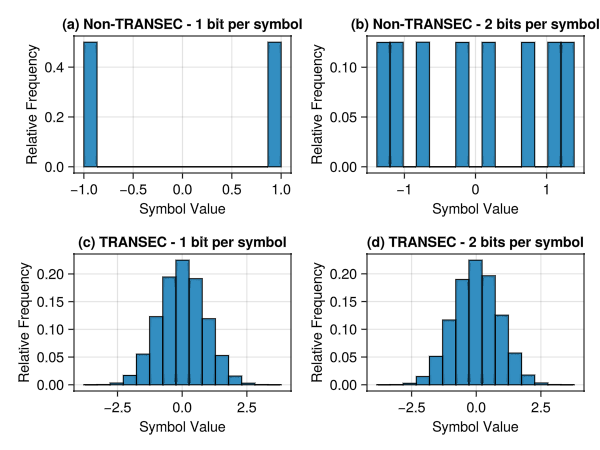}
    \caption{Histograms}
    \label{fig:combined_histogram}
\end{figure}

The histogram of the serialized symbols is shown in the top row (non-TRANSEC) of Fig. \ref{fig:combined_histogram} (a, b). The 1-bit-per-symbol case shows that the symbols are distributed as BPSK, i.e. +1 or -1. For the 2-bit-per-symbol case, the PSK modulator produces 16 PSK and then the in-phase and quadrature-phase components are serialized to produce the histogram of symbols shown. 16-PSK will be serialized into independent in-phase and quadrature-phase components thus rendering the 4-bit-per-symbol PSK into a 2-bit-per-symbol real-valued modulating signal.

The serialized symbols are then upsampled and pulse shaped. The pulse shape for this work is the root-raised cosine (RRC). The RRC pulse shaped modulating signal is then scaled by the modulation index, indicated as a gain-triangle in Fig. \ref{fig:tx_sys_diagram}. The modulation index is 0.5 (MSK).

\subsection{Adding TRANSEC to the Transmitter}
Let the serialized PSK symbols be denoted as $s[k]$, where $k$ is an integer representing the symbol index. Each symbol $s[k]$ follows the distribution of the constellation of the signal $s$. 

The PSK symbols are gathered into frames, represented as $\mathbf{s}[m]$ in \eqref{eq:def_s_block}, where $L$ is the length of the frame and $m$ is an integer representing the frame number.
\begin{equation}
\label{eq:def_s_block}
\mathbf{s}[m] = \{s[mL]; s[mL-1]; ... s[mL-L+1]\}
\end{equation}

The unitary spreader, denoted with a $U$, operates on those frames of PSK symbols $\mathbf{s}[m]$. The unitary spreader is a random unitary matrix. The elements of the unitary matrix change after each frame. This is indicated by giving $U$ a subscript $m$, representing the frame number, as shown in \eqref{eq:def_u_trans}. The output of this unitary spreading transformation will be denoted in TRANSEC frames $\mathbf{u}[m]$. 
\begin{equation}
\label{eq:def_u_trans}
\mathbf{u}[m] = U_{m}\mathbf{s}[m]
\end{equation}

Each TRANSEC frame $\mathbf{u}[m]$ has the same energy as its source symbols $\mathbf{s}[m]$ as shown in \eqref{eq:def_u_l2}. Unitary spreading preserves energy, and that distinguishes it from simply mixing the symbols with a Gaussian sequence. Because of this capability, unitary de-spreading will faithfully reconstruct the source constellation of $s$ in noise.

\begin{equation}
\label{eq:def_u_l2}
\sum_{k=mL-L+1}^{mL}\|u[k]\|^{2} = \sum_{k=mL-L+1}^{mL}\|s[k]\|^{2} 
\end{equation}

After the unitary spreading transform, the TRANSEC frames $\mathbf{u}[m]$ are serialized to now Gaussian-distributed TRANSEC symbols $u[k]$. The TRANSEC symbols $u[k]$ will be asymptotically Gaussian-distributed regardless of the underlying distribution of the symbols $s[k]$. 

When TRANSEC is enabled, the serialized symbols follow the distribution shown in the bottom row of Fig. \ref{fig:combined_histogram} (c, d). Note that the histograms for the TRANSEC 1-bit-per-symbol and 2-bit-per-symbol cases are identical. This demonstrates that the underlying modulation order has been hidden. Both cases now follow a Gaussian distribution. This is because the symbols are independent and identically distributed random variables and the output of the unitary spreader is a set of independent weighted sums of those random variables.

The TRANSEC blocks $\mathbf{u}[m]$ and TRANSEC symbols $u[k]$ follow the relation shown for $\mathbf{s}[m]$ and $s[k]$ in \eqref{eq:def_s_block}. 

The transmitter will then serialize the in-phase and quadrature-phase components of the TRANSEC symbols $u[k]$. These serialized TRANSEC symbols will then be real-valued. The serialized TRANSEC symbols are upsampled and pulse-shaped by the same RRC shape used in the non-TRANSEC case. The TRANSEC pulse-shaped symbols are normalized by the mean absolute value of their distribution, and then scaled by the modulation index as in the non-TRANSEC case. The TRANSEC modulation signal is also scaled by an additional factor, shown in Table \ref{tab:occupied_bw}. This scale affects BER and Occupied Bandwidth.

The resulting pulse-shaped TRANSEC signal drives the NCO. Crucially, since the information is encoded in the carrier phase (CPM), the output remains constant-envelope (0 dB PAPR). This ensures the enhancement is compatible with high-efficiency tactical amplifiers, avoiding the power back-off required for typical high-PAPR Gaussian waveforms. Unitary spreading preserves the symbol rate and avoids chip-based bandwidth expansion. However, the shift to Gaussian modulating statistics reshapes the non-linear CPM power spectral density, a characteristic analyzed in Section \ref{sec:spectra}.

\section{Transmit Spectra and Occupied Bandwidth}
\label{sec:spectra}
Using this transmitter system model, several CPM signals were generated: non-TRANSEC 1-bit-per-symbol (RRC MSK), non-TRANSEC 2-bit-per-symbol (16-ary PSK/FM), TRANSEC 1-bit-per-symbol, and TRANSEC 2-bit-per-symbol. These modulated signals will be analyzed throughout this work. The proposed TRANSEC enhancement applies a unitary spreading transform to CPM symbols gathered into frames of length $L=16$ at a symbol rate of 16,000 symbols/sec. These specific parameters were utilized for all subsequent testing and analysis without loss of generality.

\begin{figure}[!htbp]
    \centering
    \includegraphics[width=1\linewidth]{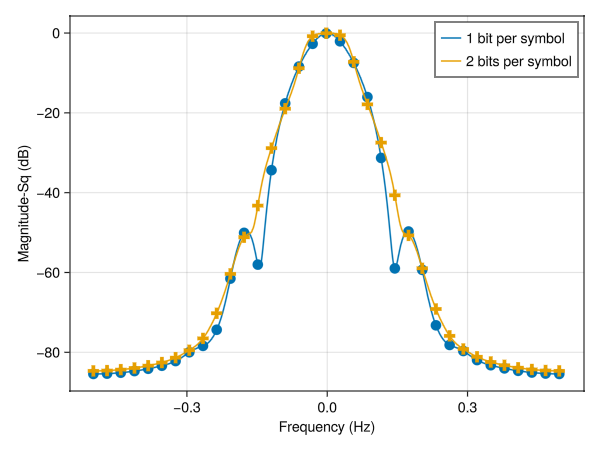}
    \caption{Non-TRANSEC Spectra}
    \label{fig:spectra}
\end{figure}

\begin{figure}[!htbp]
    \centering
    \includegraphics[width=1\linewidth]{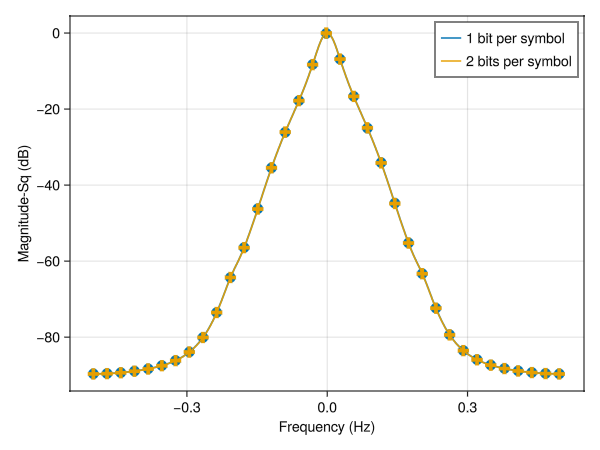}
    \caption{TRANSEC Spectra}
    \label{fig:transec_spectra}
\end{figure}

The spectra of the two non-TRANSEC CPM signals are shown in Fig. \ref{fig:spectra}. The spectrum of the 1-bit-per-symbol waveform clearly differs from the spectrum of the 2-bit-per-symbol waveform. By contrast, the spectra of the two TRANSEC CPM signals are shown in Fig. \ref{fig:transec_spectra}. The spectra are superimposed showing that the spectra of the TRANSEC enhanced waveforms are nearly identical, making it difficult if not impossible to deduce the modulation order of the signal from the spectra alone.

The 99\% occupied bandwidth of the non-TRANSEC waveform and the TRANSEC waveform with various scale factors is shown in Table \ref{tab:occupied_bw}.
\begin{table}[!htbp]
    \caption{Occupied Bandwidth}
    \label{tab:occupied_bw}
    \centering
    \begin{tabular}{ |c|c|c| }
        \hline
        Bits per Symbol & 1 & 2\\
        \hline
        Non-TRANSEC & 17.6 kHz & 19.7 kHz\\
        \hline
        TRANSEC and Scale = 0.7 & 19.6 kHz & 19.6 kHz\\
        \hline
        TRANSEC and  Scale = 0.9 & 23.7 kHz & 23.7 kHz\\
        \hline
    \end{tabular}
\end{table}

Note that when the non-TRANSEC waveform changes from 1-bit-per-symbol to 2-bits-per-symbol, the bandwidth increases. This is not so in the TRANSEC cases. The bandwidth of the TRANSEC signals is not a function of the underlying modulation. This denies a feature for the adversary to observe and enhances the anonymity of the transmitter.

The scaling factor of the TRANSEC signals acts as a spectral control mechanism, limiting the peak frequency deviation of the CPM carrier. As will be seen in section \ref{sec:results}, this allows the signal to fit specific frequency masks, trading a predictable BER penalty for a reduced spectral footprint.

\subsection{Receiver}
\begin{figure}[!htbp]
    \centering
    \includegraphics[width=1\linewidth]{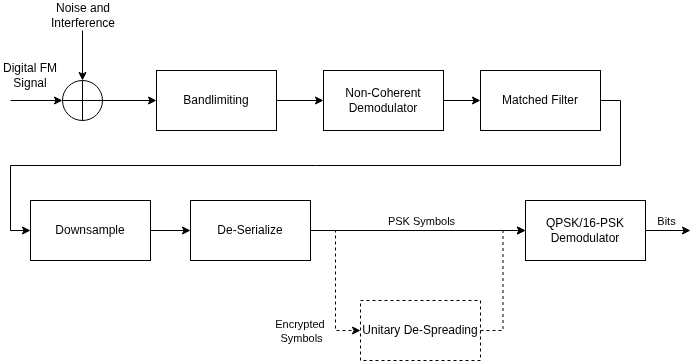}
    \caption{Receiver System Diagram}
    \label{fig:rx_sys_diagram}
\end{figure}

A diagram of the receiver is shown in Fig. \ref{fig:rx_sys_diagram}. The received signal is a complex-valued baseband signal. The modulation is CPM. Noise and Interference are added to the signal. Noise represents noise in the receiver as a function of $E_s/N_0$. The noisy signal is bandlimited before passing into a noncoherent FM demodulator. The noncoherent FM Demodulator will also flag samples that have been saturated. After demodulation, the output is matched-filtered and downsampled to one sample per symbol. The real-valued downsampled signal is then de-serialized such that adjacent pairs of samples become complex-valued samples. If TRANSEC is not enabled, the now complex-valued signal will go to PSK demodulation. If TRANSEC is enabled, the complex-valued signal will go through unitary de-spreading, and then PSK demodulation.

\subsection{Adding TRANSEC to the Receiver}
The received signal $r[k]$ is composed of the TRANSEC signal $u[k]$ and additive white Gaussian noise (AWGN) $n[k]$ as shown in \eqref{eq:def_r}. 
\begin{equation}
\label{eq:def_r}
    r[k] = u[k] + n[k]
\end{equation}

The received signal samples are grouped into frames $\mathbf{r}[m]$ as described for $s$ in \eqref{eq:def_s_block}. These frames of samples of $r$ are composed of samples of $u$ and $n$ as shown in \eqref{eq:def_r_block}.

\begin{equation}
\label{eq:def_r_block}
\mathbf{r}[m] = \mathbf{u}[m] + \mathbf{n}[m]
\end{equation}

The frames $\mathbf{r}[m]$ are then de-spread by a conjugate-transpose of the unitary spreading matrix as shown in \eqref{eq:def_u_decrypt}. 
\begin{equation}
\label{eq:def_u_decrypt}
\mathbf{s}[m] + \mathbf{w}[m] = U_{m}^{H}\mathbf{r}[m]
\end{equation}

The conjugate transpose of $U$ has a block index $m$ as was used in the spreading process. The receiver shares a secret key with the transmitter from which $U_{m}$ is derived. There is a noise term $w$ generated by unitary de-spreading the original AWGN signal $n$. Note that this unitary de-spreading causes no amplification of noise. Because this process is linear, the signal $s$ is faithfully recovered and has all of its original signal energy. The PSK demodulator must still contend with the noise term $w$, this process does not improve $E_s/N_0$. BER curves will be analyzed in the following section.


%
%
%

\section{Results}
\label{sec:results}
\subsection{Cyclostationary Analysis}

\begin{figure}[!htbp]
    \centering
    \includegraphics[width=1\linewidth]{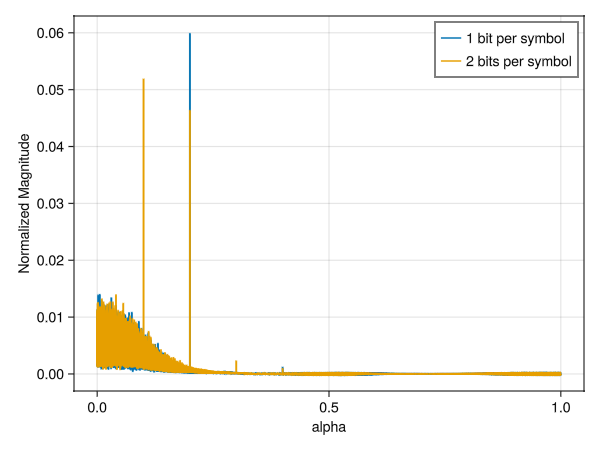}
    \caption{Non-TRANSEC Alpha Profile Comparison}
    \label{fig:caf_comparison}
\end{figure}

\begin{figure}[!htbp]
    \centering
    \includegraphics[width=1\linewidth]{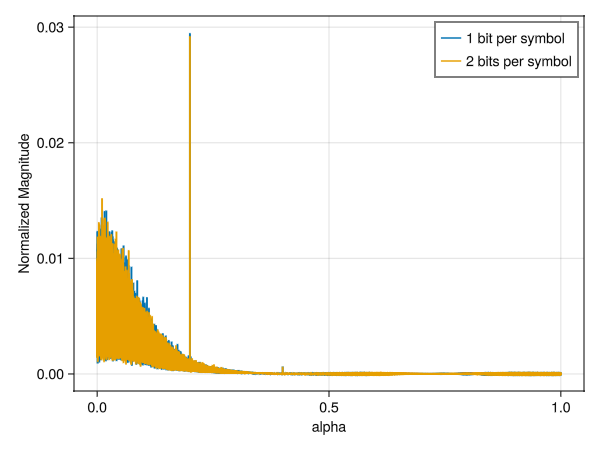}
    \caption{TRANSEC Alpha Profile Comparison, Scale = 0.7}
    \label{fig:transec_caf_comparison}
\end{figure}

Fig. \ref{fig:caf_comparison} shows the alpha profiles of the two non-TRANSEC CPM signals overlaid on each other. The alpha profile of the 1-bit-per-symbol non-TRANSEC CPM signal contains a single large baud line at $\alpha = 0.2$, whereas the alpha profile of the 2-bits-per-symbol non-TRANSEC CPM signal contains two large baud lines at $\alpha = 0.2$ and $\alpha = 0.1$. The difference between the two is immediately recognizable and any adversary can determine which of the two is in use. Fig. \ref{fig:transec_caf_comparison} shows nearly identical alpha profiles for both of the TRANSEC CPM signals, denying that feature to the adversary. The values of the maxima of the alpha profile plots are shown in Table \ref{tab:alpha_max}. The TRANSEC waveform using a scaling factor of 0.9 is not featured in a figure to save space, but the maximum at $\alpha = 0.2$ is provided. The values show the TRANSEC enhancement attenuating the feature.

\begin{table}[!htbp]
    \caption{Alpha Profile Maxima}
    \label{tab:alpha_max}
    \centering
    \begin{tabular}{ |c|c|c| }
        \hline
        Number of Bits per Symbol & 1 & 2 \\
        \hline
        Non-TRANSEC & 0.0600 & 0.0519\footnotemark[1]\\
        \hline
        TRANSEC and Scale = 0.7 & 0.0295 & 0.0292\\
        \hline
        TRANSEC and  Scale = 0.9 & 0.0381 & 0.0378\\
        \hline
    \end{tabular}

    
  \vspace{1ex} 
  {\footnotesize \raggedright \footnotemark[1]Maximal value at $\alpha=0.1$; all other maxima occur at $\alpha=0.2$\par}    
\end{table}

\subsection{BER in AWGN}
\begin{figure}[!htbp]
    \centering
    \includegraphics[width=1\linewidth]{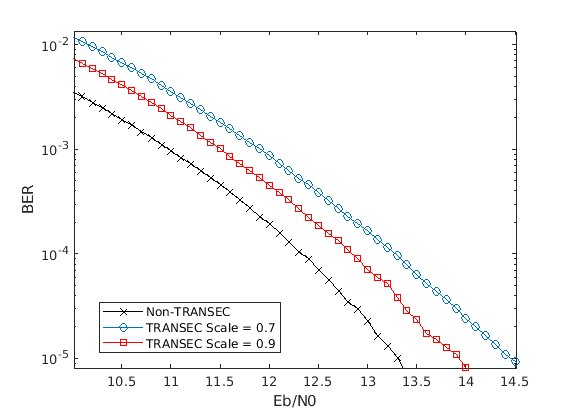}
    \caption{BER Curve 1-bit-per-symbol}
    \label{fig:ber_curve_2_scale_0.5}
\end{figure}

The BER results in AWGN for the 1-bit-per-symbol case are shown in Fig. \ref{fig:ber_curve_2_scale_0.5}. The BER performance of the TRANSEC waveform with scaling of 0.7 is just over 1 dB worse than the non-TRANSEC waveform. This BER performance improves with the 0.9 scaling, but at the expense of bandwidth as detailed in Table \ref{tab:occupied_bw}.

\begin{figure}[!htbp]
    \centering
    \includegraphics[width=1\linewidth]{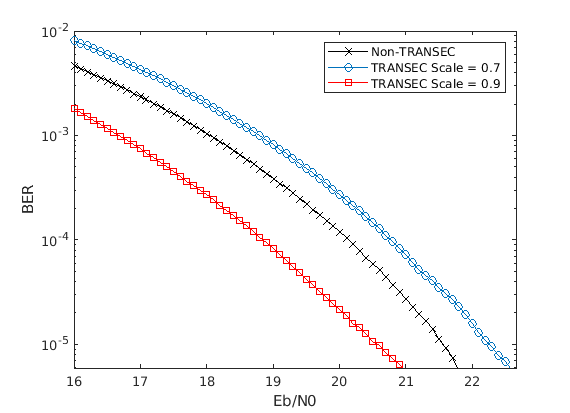}
    \caption{BER Curve 2-bits-per-symbol}
    \label{fig:ber_curve_4_scale_0.5}
\end{figure}

The BER results in AWGN for the 2-bits-per-symbol case are shown in Fig. \ref{fig:ber_curve_4_scale_0.5}. The TRANSEC waveform with scaling of 0.7 is just under 1 dB worse than the non-TRANSEC waveform and has about the same bandwidth. The TRANSEC waveform with 0.9 scaling uses more bandwidth and outperforms the non-TRANSEC waveform by a little over 1 dB.

\subsection{BER in AWGN and Interference}
In this experiment, a strong interferer was added to the receiver model. The interference was modeled as a broadband blocking jammer that saturated the front-end of the receiver. This saturation results in the blanking of the post-demodulated data. Because the interference saturates the receiver, the blanked samples are known. From these indices a diagonal blanking matrix $B_m$ can be determined. Note that $B_m$ will change with every m-th frame. 

For this experiment, we introduce a convex optimization option for the TRANSEC waveform. The convex optimization works as follows:
Consider the ML criterion given in \eqref{eq:ml_func}, where $\mathcal{S}$ is the set of symbols.
\begin{equation}
\label{eq:ml_func}
\mathbf{\hat{s}}[m] = \argmin_{\mathbf{s}[m]} \|B_m U_m \mathbf{s}[m] - \mathbf{r}[m] \|_2^2 \quad \text{subject to} \quad s \in \mathcal{S}
\end{equation}

Box-relaxation changes the constraints to any values within the box defined by the QPSK constellation, as shown in \eqref{eq:cost_func}
\begin{multline}
\label{eq:cost_func}
\mathbf{\hat{s}}[m] = \argmin_{\mathbf{s}[m]} \|B_m U_m \mathbf{s}[m] - \mathbf{r}[m] \|_2^2 \\
\text{subject to} \quad |Re(s)| < 1 \quad and \quad  |Im(s)| < 1
\end{multline}

Taking the derivative of \eqref{eq:cost_func} with respect to $\mathbf{s}[m]$ yields an update function of 
\begin{equation}
\label{eq:update_func}
2 (B_m U_m)^H (B_m U_m \mathbf{s}[m] - \mathbf{r}[m])
\end{equation}

\begin{figure}[!htbp]
    \centering
    \includegraphics[width=1\linewidth]{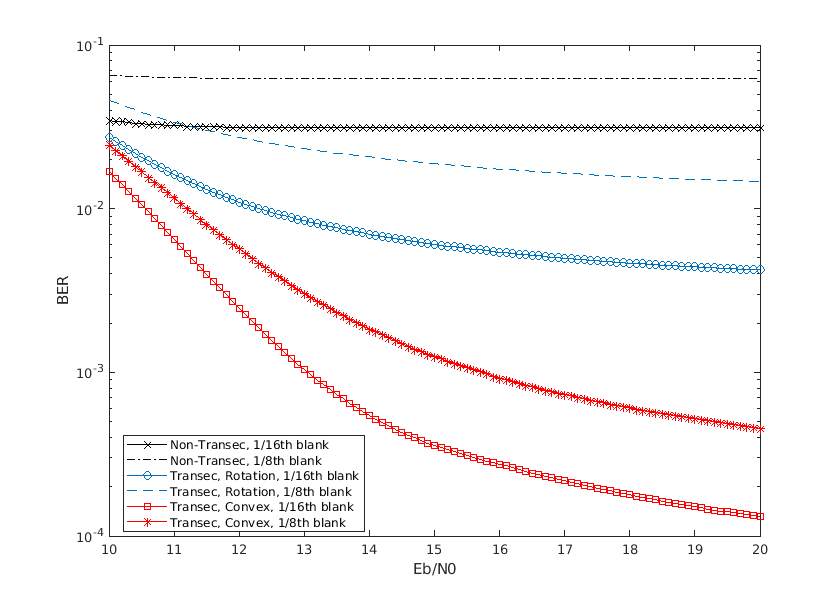}
    \caption{BER Curves with Erasures}
    \label{fig:ber_curve_erasures}
\end{figure}

Using 1-bit-per-symbol, we ran the non-TRANSEC receiver, the TRANSEC receiver with only de-spreading, and the TRANSEC receiver with convex optimization. The results are shown in Fig. \ref{fig:ber_curve_erasures}. The three waveforms were tested with 1/16th of samples blanked and 1/8th of samples blanked. The non-TRANSEC performs the worst with error floors of 0.03125 and 0.0625, respectively. The de-spreading-only method also hits an early error floor. The convex optimization method performs the best lowering the error floor by multiple orders of magnitude. The TRANSEC waveform used a scaling factor of 0.7. The recovery performance of this technique does not plateau at 12.5\% blanking; for example, it recovers the signal at 25\% blanking with a BER of 0.0059 at 20 dB $E_b/N_0$.

\section{Conclusion}
This work has demonstrated the efficacy of a unitary spreading-based TRANSEC enhancement to protect the physical-layer characteristics of CPM waveforms. Our findings reveal that the non-TRANSEC CPM waveforms exhibit highly distinct symbol histograms, spectral profiles, and prominent baud lines in their cyclic autocorrelation functions. The introduction of this TRANSEC enhancement mitigates these signatures. We have shown that we can frustrate cyclostationary feature extraction thus enhancing LPI. We have shown that spectral features give no indication of the modulation type underneath the unitary spreading. 

We demonstrated that the TRANSEC enhancement provides resilience in the presence of jamming. We explored two different methods of recovering the unitary spread symbols, simple derotation and convex optimization. The optimization method was able to recover symbols with a low bit error floor. These experiments were run with uncoded data. 

This TRANSEC enhancement can augment existing FEC schemes. Adding FEC will only improve the performance. This enhancement can be integrated with other spreading techniques. We leave integration of this enhancement with FEC and spreading schemes to future work. 

An evaluation of system performance highlights a direct trade-off between operational bandwidth and bit error rate. Although the non-TRANSEC waveforms inherently achieve a lower BER than their encrypted counterparts at equivalent bandwidths, that loss is small. 

Ultimately, this TRANSEC enhancement provides a highly secure, adaptive solution for tactical communications, successfully denying electronic warfare systems the cyclostationary features needed to exploit and track deployed forces, and recovering data from a broadband blocking jammer.

\FloatBarrier
\bibliographystyle{IEEEtran}
\bibliography{transecfsk.bib}

\end{document}